\documentclass[twocolumn,epjc3]{svjour3}

\usepackage{subfigure}
\usepackage{graphicx}  
\usepackage{dcolumn}   
\usepackage{bm}        
\usepackage{amssymb}   

\usepackage[hyperindex,breaklinks]{hyperref}
\usepackage{url}
\usepackage{footnote}

\journalname{Eur. Phys. J. C}

\def\MET{E^{\rm miss}_{\rm T}}
\def\antibar#1{\ensuremath{#1\bar{#1}}}
\def\ttbar{\antibar{t}}

\def\mmjj{\mu\mu jj}
\def\mnjj{\mu\nu jj}

\usepackage{preprintcover}  

\PreprintCoverPaperTitle{Search for second generation scalar leptoquarks in \\ $pp$ collisions at $\sqrt{s}=7$ TeV with the ATLAS detector}

\PreprintIdNumber{CERN-PH-EP-2012-056}

\PreprintCoverAbstract{The results of a search for the production of second generation scalar leptoquarks are 
presented for final states consisting of either two muons and at least 
two jets or a muon plus missing transverse momentum and at least two jets. A total of 
1.03~fb$^{-1}$ integrated luminosity of proton-proton collision data produced by the Large Hadron Collider at $\sqrt{s}=7$~TeV 
and recorded by the ATLAS detector is used for the search.
The event yields in the signal regions are found to be consistent with the Standard Model background 
expectations. The production of second generation leptoquarks is excluded for a leptoquark mass $m_{\rm LQ}<$ 594 (685)~GeV 
at 95\% confidence level, for a branching ratio of 0.5 (1.0) for leptoquark decay to a muon and a quark. }

\PreprintJournalName{European Physical Journal C}

\begin{document}

\title{Search for second generation scalar leptoquarks in \\ $pp$ collisions at $\sqrt{s}=7$ TeV with the ATLAS detector}
\titlerunning{Search for second generation scalar leptoquarks in $pp$ collisions at $\sqrt{s}=7$ TeV with the ATLAS detector}

\author{The ATLAS Collaboration}
\institute{}
\date{August 16, 2012}

\maketitle

\begin{abstract}
The results of a search for the production of second generation scalar leptoquarks are 
presented for final states consisting of either two muons and at least 
two jets or a muon plus missing transverse momentum and at least two jets. A total of 
1.03~fb$^{-1}$ integrated luminosity of proton-proton collision data produced by the Large Hadron Collider at $\sqrt{s}=7$~TeV 
and recorded by the ATLAS detector is used for the search.
The event yields in the signal regions are found to be consistent with the Standard Model background 
expectations. The production of second generation leptoquarks is excluded for a leptoquark mass $m_{\rm LQ}<$ 594 (685)~GeV 
at 95\% confidence level, for a branching ratio of 0.5 (1.0) for leptoquark decay to a muon and a quark. 
\end{abstract}


\section{Introduction}
\label{sec-intro}

The remarkable similarities between quarks and leptons in the Standard Model (SM) lead
to the supposition that there could be a fundamental relationship between them at a sufficiently 
high 
energy scale, manifested by the existence of leptoquarks (LQ)~\cite{lqpapers}. LQs are 
hypothetical particles which carry both baryon and lepton number and have fractional electrical 
charge. The present search is performed within the minimal Buchm\"{u}ller-R\"{u}ckl-Wyler  
model~\cite{mBRW}, where LQs are restricted to couple to quarks and leptons of one generation.
In this model, LQs are required to have pure chiral couplings to SM fermions in order 
to avoid inducing four-fermion interactions that would cause flavour-changing neutral currents 
and lepton family-number violations. At the Large Hadron Collider (LHC), scalar LQs can be produced either in pairs or 
singly. Single LQ production involves the unknown $\lambda_{LQ-\ell-q}$ coupling, 
while pair production of scalar LQs occurs mostly via gluon-gluon fusion, dominant for 
$m_{\rm LQ}\lesssim 1$~TeV, and $q\overline{q}$-annihilation, dominant at larger masses. 
Both pair-production modes involve only the strong coupling constant, and therefore all model 
dependence is contained in the assumed LQ mass $m_{\rm LQ}$ and the branching ratio $\beta$ for LQ decay 
to a charged lepton and a quark
\footnote{The $\lambda_{LQ-\ell-q}$ coupling determines the LQ
lifetime and width~\cite{b-comphep}. For LQ masses considered here, 200~GeV$\leq m_{\rm LQ} \leq$700~GeV, couplings
greater than $e \times 10^{-6}$, with $e=\sqrt{4\pi\alpha}$ the electron charge, and $\alpha(M_{\rm Z})=1/128$, 
correspond to decay lengths less than roughly 1~mm. In addition, to be insensitive to the coupling, 
the width cannot be larger than the experimental resolution of a few GeV. This sets the approximate sensitivity 
to the unknown coupling strength.}. LQs can also decay to a neutrino and a quark; in this case, the 
branching ratio is $1-\beta$. Pair production of scalar LQs at the LHC
has been calculated at next-to-leading order (NLO)~\cite{b-kramer}. 

The results presented in this paper are an update of the previous ATLAS search for second generation LQs~\cite{LQ_LHC2} and 
extend the bounds arising from previous direct searches performed by CMS~\cite{LQ_LHC1}, ATLAS~\cite{LQ_LHC2},
D0~\cite{LQ_Fermilab} and OPAL~\cite{LQ_LEP}. 
A total integrated luminosity of 1.03~fb$^{-1}$ of proton-proton collision data at a centre of mass
energy $\sqrt{s}=7$~TeV, collected with the ATLAS detector from March through 
July 2011, is used for the search. The final states arising 
from leptoquark pairs decaying into two muons and two quarks ($\mmjj$), 
or into a muon, a neutrino and two quarks ($\mnjj$), are considered. These result in experimental 
signatures of either two high transverse momentum ($p_{\rm T}$) muons and two high $p_{\rm T}$ jets,
or one high $p_{\rm T}$ muon, missing transverse momentum, and two high $p_{\rm T}$ jets. 

Analyses for both dimuon and single muon final states start with the selection of event samples with
large signal acceptance. Since background cross sections are several orders of magnitude larger 
than the signal cross sections, these samples are dominated by the major backgrounds: 
$Z$+jets and $\ttbar$ in the $\mmjj$ case, and $W$+jets and $\ttbar$ for the $\mnjj$
case. Further selection requirements are then applied to these samples to define control regions used 
to determine the normalization of the aforementioned backgrounds. The determination of the
multi-jet background is performed in a fully data-driven approach, and the smaller diboson and single top-quark
backgrounds are estimated using Monte Carlo (MC) simulations.

After all background contributions are determined, variables selected to enhance 
the discrimination between signal and background are combined into a 
log likelihood ratio, which is used to search for an excess of events over the SM background
prediction. The searches are performed independently for each final state. The results 
are then combined and interpreted as lower bounds on the LQ mass for different $\beta$ hypotheses.

\section{The ATLAS detector}
\label{sec-atlas}

The ATLAS detector~\cite{atlas} is a multi-purpose detector with a forward-backward symmetric 
cylindrical geometry
and nearly 4$\pi$ coverage in solid angle
\footnote{ATLAS uses a right-handed coordinate 
system with its origin at the nominal interaction point and the \textit {z}-axis along the beam pipe. 
Cylindrical 
coordinates (\textit r,$\phi$) are used in the transverse plane, with $\phi$ the azimuthal angle 
around the beam pipe. The pseudorapidity $\eta$ is defined in terms of the polar angle $\theta$ by 
$\eta$ = --ln tan($\theta$/2).}. 

The three major sub-components of ATLAS are the tracking detectors, the calorimeters and the 
muon spectrometer. Charged particle tracks and vertices are reconstructed with 
silicon-based tracking detectors that cover $|\eta| <$ 2.5 and a transition radiation tracker 
extending to $|\eta| <$ 2.0. 
The inner tracking system is immersed in a homogeneous 2~T axial magnetic field provided by a solenoid. 
Electron, photon, and jet energies are measured in the calorimeters. The calorimeter 
system is segmented into a central barrel and two endcaps, collectively covering the pseudorapidity 
range of $|\eta| <$ 4.9. A liquid-argon (LAr) electromagnetic calorimeter covers the range 
$|\eta|<3.2$ and an iron-scintillator tile hadronic calorimeter covers the range $|\eta|<1.7$. 
Endcap and forward LAr calorimeters provide both electromagnetic and hadronic measurements
and cover the region $1.5<|\eta|<4.9$. 

Surrounding 
the calorimeters, a muon spectrometer~\cite{atlas} with air-core toroids, a system of precision tracking chambers, 
and detectors with triggering capabilities provides muon identification and precise momentum measurements. The muon spectrometer is based on three large superconducting toroids with coils arranged in an eight-fold symmetry around the calorimeters, covering a range of $|\eta|<$ 2.7. Over most of the $\eta$ range, precision measurements of the track coordinates in the principal bending direction of the magnetic field are provided by Monitored Drift Tubes (MDTs). At large pseudorapidities (2.0 $< |\eta| < $ 2.7), Cathode Strip Chambers (CSCs) with higher granularity are used in the innermost station. 

A three-level trigger system selects events to be recorded for offline analysis. The muon trigger detectors consist of Resistive Plate Chambers (RPCs) in the barrel ($|\eta|<$ 1.05) and Thin Gap Chambers (TGCs) in the end-cap regions (1.05 $< |\eta| <$ 2.4), with a small overlap in the $|\eta|=$ 1.05 region. The data considered
in this analysis are selected from events containing at least one muon with the transverse momentum
determined by the trigger system satisfying $p_{\rm T}>18$~GeV.

\section{Simulated samples}
\label{sec-mc}

Simulated event samples are used to determine all signal and some of the background yields. 
Signal samples for LQ masses between 200 GeV and 1000~GeV are simulated with 
PYTHIA 6.4.25~\cite{b-pythia}. NLO cross sections as determined in Ref.~\cite{b-kramer},
using CTEQ6.6~\cite{b-cteq66} parton distribution functions (PDFs), are used to normalize the samples 
at each mass point. 
 
Samples of $W$ and $Z/\gamma^{\star}$ production in association with $n$ partons (where $n$ can be 0, 1, 2, 3, 4 
and 5 or more) are simulated 
with the ALPGEN~\cite{b-alpgen} generator interfaced to HERWIG~\cite{b-herwig} and JIMMY~\cite{b-jimmy} 
to model parton showers and multiple parton interactions, respectively. The MLM~\cite{b-alpgen} 
parton-shower matching scheme is used to form inclusive $W/Z+$jets samples. MC@NLO \cite{b-mcnlo} 
is used to estimate the production of single top quarks and top quark pairs. A top quark mass of 172.5 GeV 
is used in the simulation. Diboson events are generated using HERWIG, and the cross sections are scaled to
NLO calculations~\cite{b-mcnlo,b-mcfm}.

All simulated events are passed through a full detector simulation based on 
GEANT4~\cite{b-geant4} and then reconstructed with 
the same software chain as the data~\cite{b-atlassim}. During the data-taking period considered 
in this search, the 
mean number of primary proton-proton interactions per bunch crossing was approximately six. 
The effect of this pile-up is taken into account in the analysis by overlaying simulated
minimum bias events onto the simulated hard-scattering events. The MC samples are then 
reweighted such that the average number of pile-up interactions matches that seen in the data.

\section{Object and Event selection}
\label{sec-objsel}

Collision events are identified by requiring at least one reconstructed primary vertex candidate with at
least three
associated tracks with $p_{\rm T, track}>0.4$~GeV. If two or more such vertices are found, 
the one with the largest sum of $p_{\rm T, track}^{\rm 2}$ is taken to be the primary vertex. 
Muons are reconstructed by matching tracks in the inner 
detector to track segments in the muon spectrometers, as described in Ref.~\cite{MCP}. In addition to the track 
quality requirements imposed for identification, the muon tracks must also satisfy
$|d_0| <$ 0.1~mm and $|z_0| <$ 5~mm, where $d_0$ and $z_0$ are the transverse and longitudinal impact 
parameters measured with respect to the primary vertex. All selected muons must have 
$p_{\rm T} >$ 30~GeV and are restricted to be within $|\eta| <$ 2.4. Muon candidates must pass the isolation 
requirement $p^{\rm cone20}_{\rm T}$/$p_{\rm T} <$ 0.2, where $p^{\rm cone20}_{\rm T}$ is the sum 
of the $p_{\rm T}$ of the tracks within $\Delta R = \sqrt {(\Delta \phi)^2 + (\Delta \eta^2)} <$ 0.2 of the muon track, excluding the muon 
$p_{\rm T}$ contribution. Selected events must have at least one
 muon identified by the trigger system within a cone $\Delta R <$ 0.1 centered on a selected muon.

Jets are reconstructed from calorimeter energy clusters using the anti-${k_t}$ algorithm~\cite{b-akt} 
with a radius parameter $\it{R}$ = 0.4. Corrections are applied in order to account for the effects 
of the non-compensating calorimeter, upstream material and other effects, by using $p_{\rm T}$ 
and $\eta$-dependent correction factors derived from simulation and validated with test-beam~\cite{b-jets-testbeam} and collision data studies~\cite{b-jetpaper}. 
After applying quality requirements based on shower shape and signal timing with respect to 
the beam crossing, the selected jets must satisfy $p_{\rm T}>30$ GeV, $|\eta| <$ 2.8 and must be 
separated from the selected candidate muons by $\Delta R \geqslant 0.4$. 
The presence of neutrinos is inferred from the missing transverse momentum $\MET$, 
defined as the magnitude of the negative vector sum of the
transverse momenta of reconstructed electrons, muons and jets, as well as calorimeter energy deposits not
associated to reconstructed objects.

Corrections to the muon trigger and reconstruction efficiencies and to the
momentum resolution are applied to the simulated events so that their kinematic
distributions match those observed in data, with an impact on the predicted number
of events of less than 2\%. These corrections are derived from samples of
$Z\rightarrow\mu\mu$ and $W\rightarrow\mu\nu$  decays~\cite{MCP}, taking into account the effects of multiple
scattering and the intrinsic resolution of the muon spectrometer~\cite{b-smear}. In order
to validate the corrections at high $p_{\rm T}$, the alignment of the muon spectrometer,
which dominates the momentum resolution for $p_{\rm T}$ larger than approximately 200~GeV, 
is derived from a sample of straight track data taken in special runs with the toroids turned off,
resulting in agreement within the considered systematic uncertainties. 

Events selected for this search are required to contain either exactly two muons
and at least two jets for the $\mmjj$ final state, or exactly one muon, at least 
two jets and $\MET>$ 30~GeV for the $\mnjj$ final state. 
In the $\mmjj$ channel, only events with $m_{\mu\mu}>40$~GeV are considered. In the $\mnjj$ 
channel, events are required to have $m_{\rm T}=\sqrt{2p^{\mu}_{\rm T}\MET(1-\cos(\Delta\phi))}>40$~GeV, 
where $\Delta\phi$ is the angle between the muon and the $\MET$ direction in the plane perpendicular to 
the beam. Events with identified electrons as defined in Ref.~\cite{b-lq1}, with $p_{\rm T}>30$~GeV, and
$|\eta|<2.47$ are rejected. 
After all the selection criteria are applied the acceptance times efficiency ranges from about 
60\% (55\%) for a LQ signal of $m_{\rm LQ}=300$~GeV to 65\% (60\%) for a LQ signal of $m_{\rm LQ}=600$~GeV
for the $\mmjj$ ($\mnjj$) channel.

\begin{figure*}[htbp!]
\begin{center}
    \subfigure[]{\includegraphics[width= 0.49\textwidth]{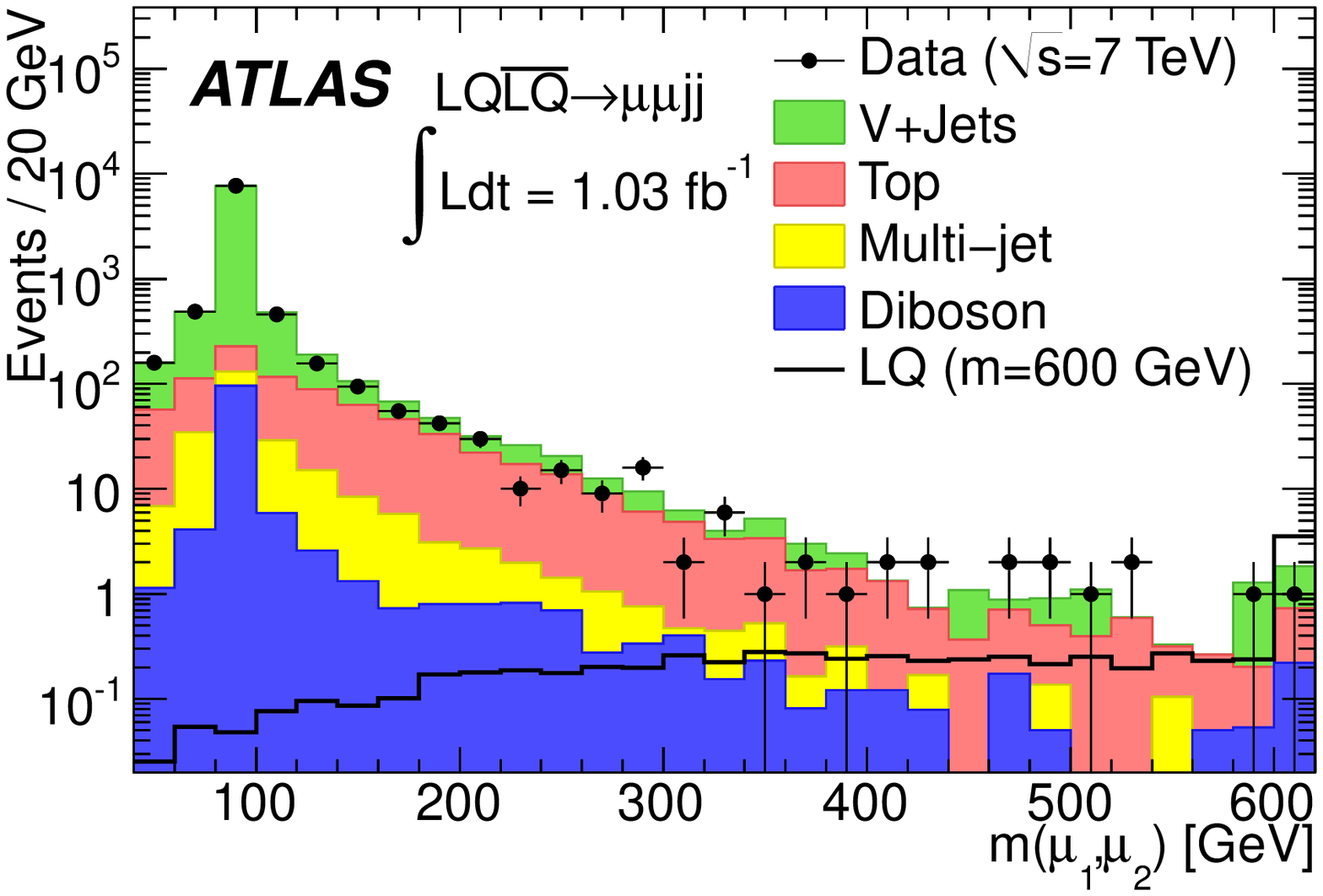}\label{f-mee}}
    \subfigure[]{\includegraphics[width= 0.49\textwidth]{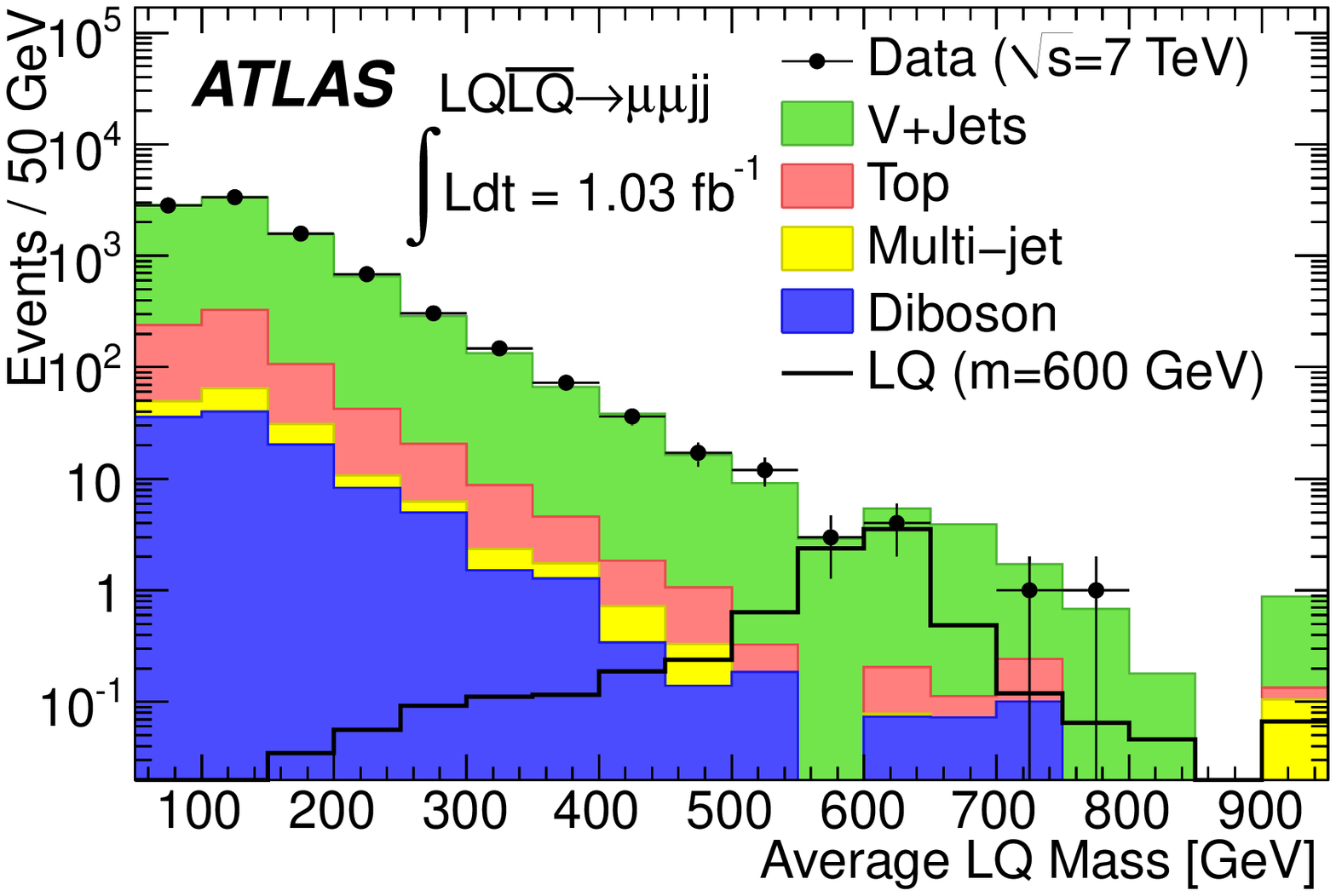}\label{f-lqm}}
    \subfigure[]{\includegraphics[width= 0.49\textwidth]{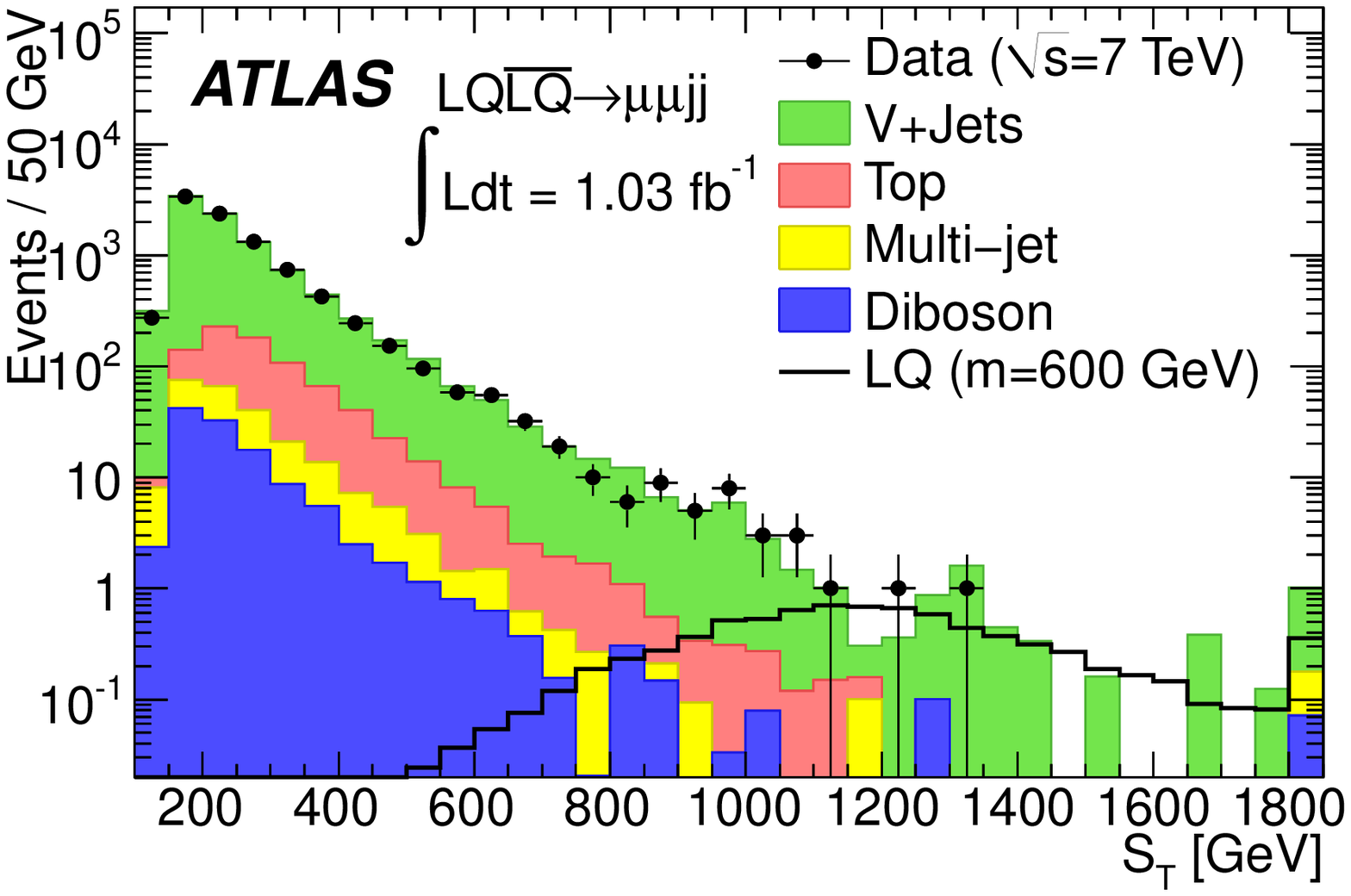}\label{f-st}}
\end{center}
    \caption{Distributions of the input $LLR$ variables for the $\mmjj$ channel for data and the SM backgrounds. 
     \subref{f-mee} Invariant mass of the two muons in the event, 
    \subref{f-lqm} Average LQ mass resulting from the best muon-jet combinations in each event, 
    and \subref{f-st} $S_{\rm T}$. The stacked distributions show the various background
    contributions, and data are indicated by the points with error bars. The 600~GeV LQ signal is
    also shown for $\beta=1.0$. 
       In all figures, the last bin contains the sum of all entries equal to and above the bin lower boundary.
\label{f-inputs-mmjj}} 
\end{figure*}

\begin{figure*}[!htbp]
\centering
  \subfigure[]{\includegraphics[width= 0.49\textwidth]{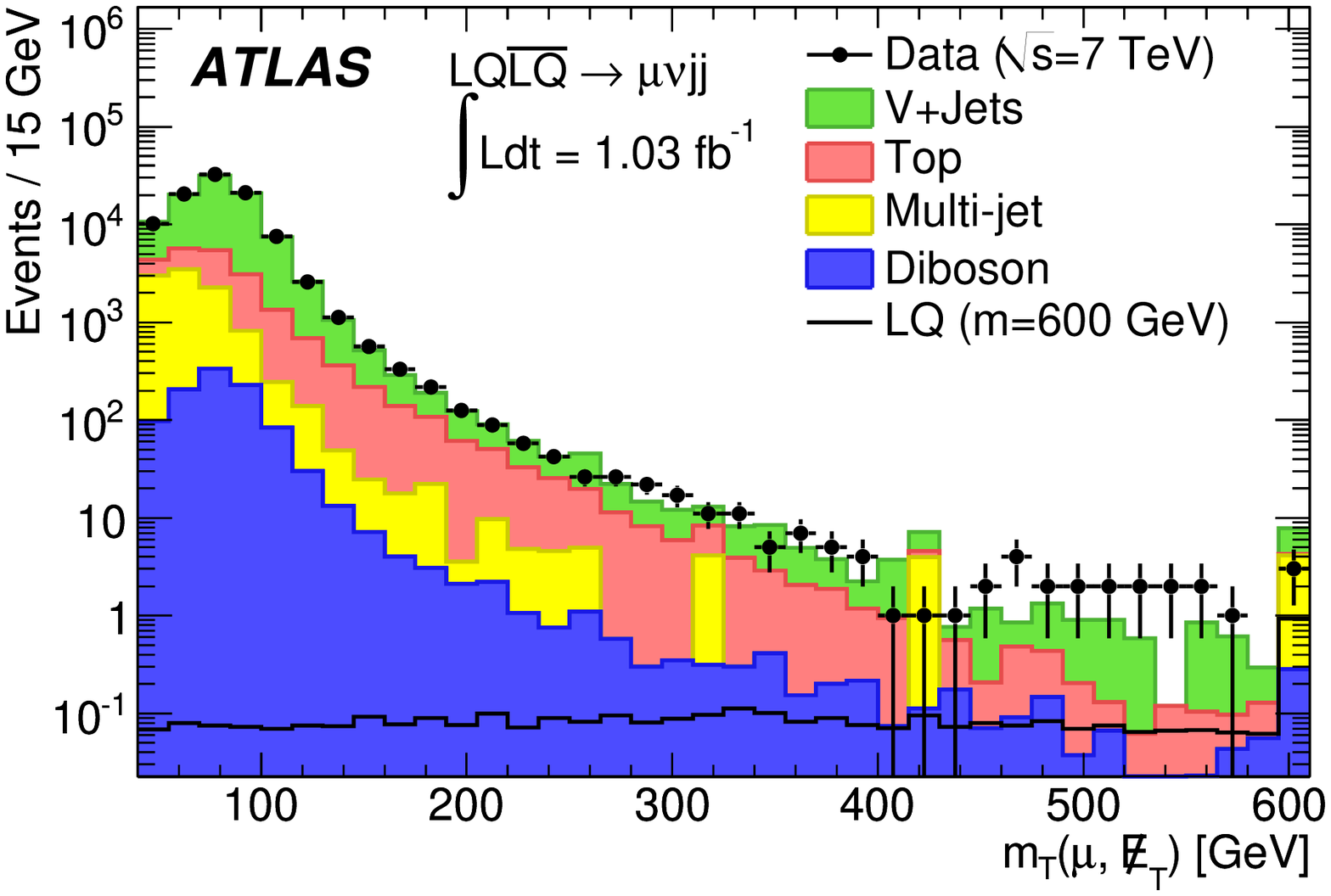}\label{f-mt}}
  \subfigure[]{\includegraphics[width= 0.49\textwidth]{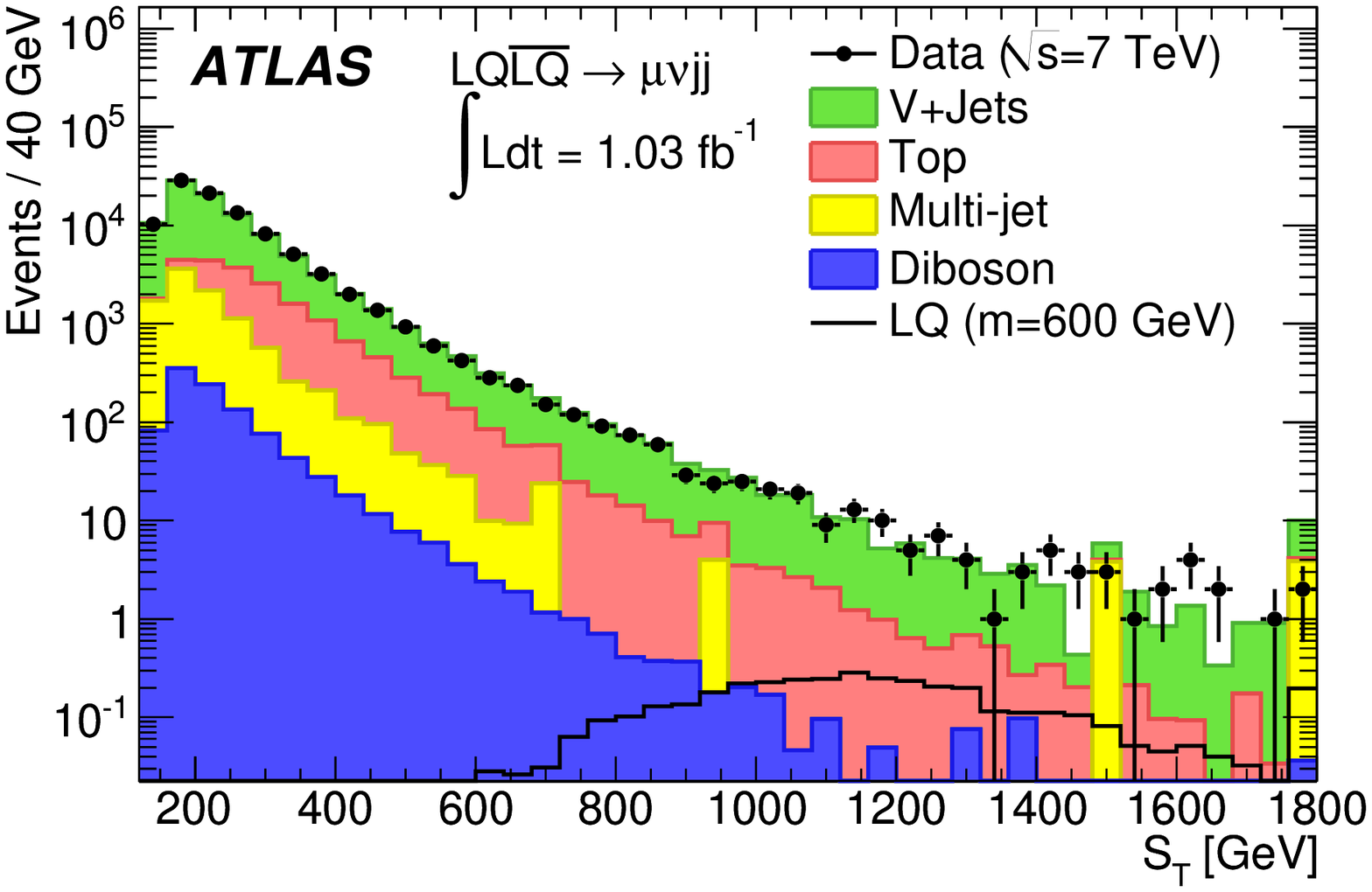}\label{f-st-mvjj}}
  \subfigure[]{\includegraphics[width= 0.49\textwidth]{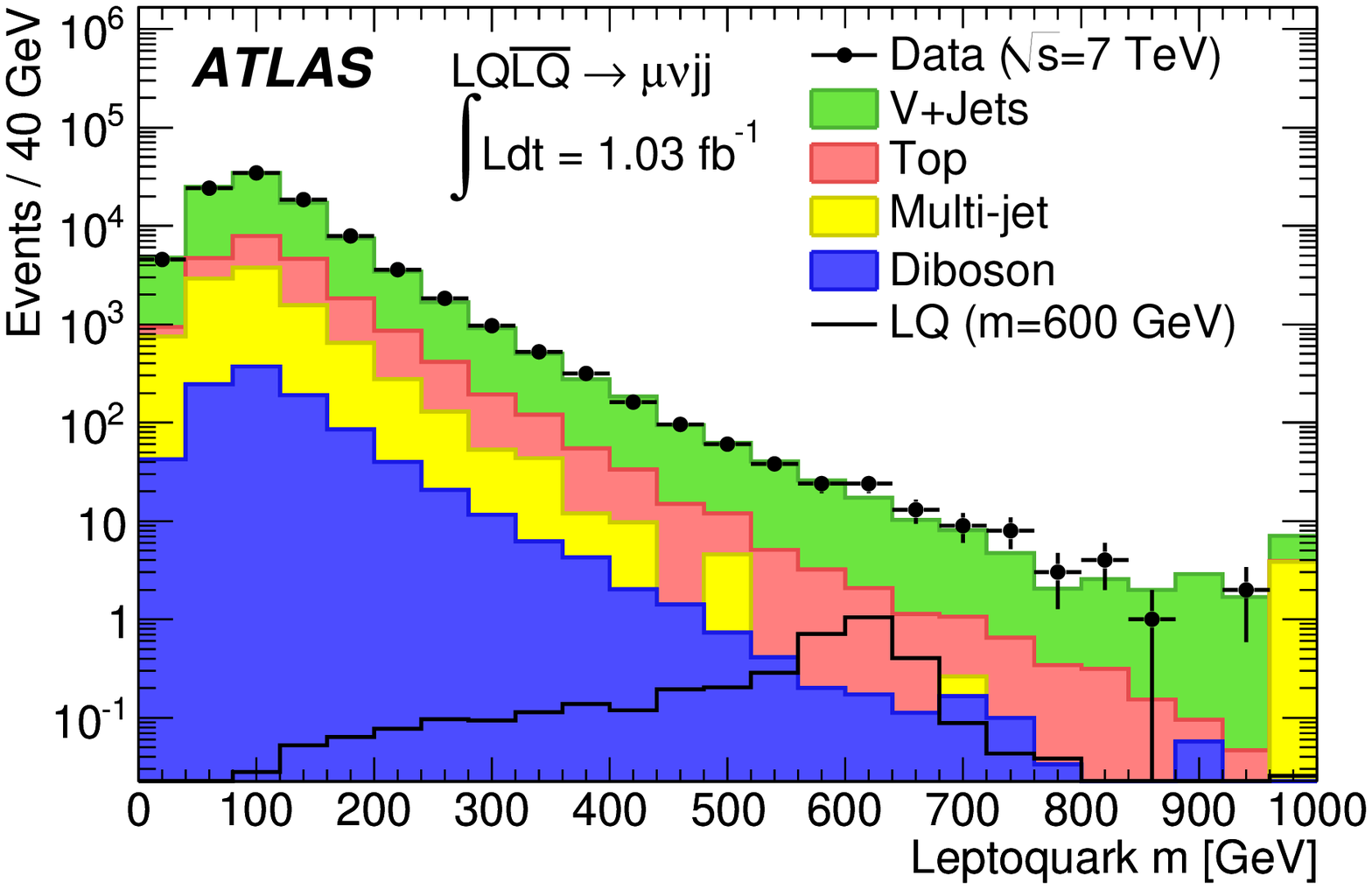}\label{f-lqm-mvjj}}
  \subfigure[]{\includegraphics[width= 0.49\textwidth]{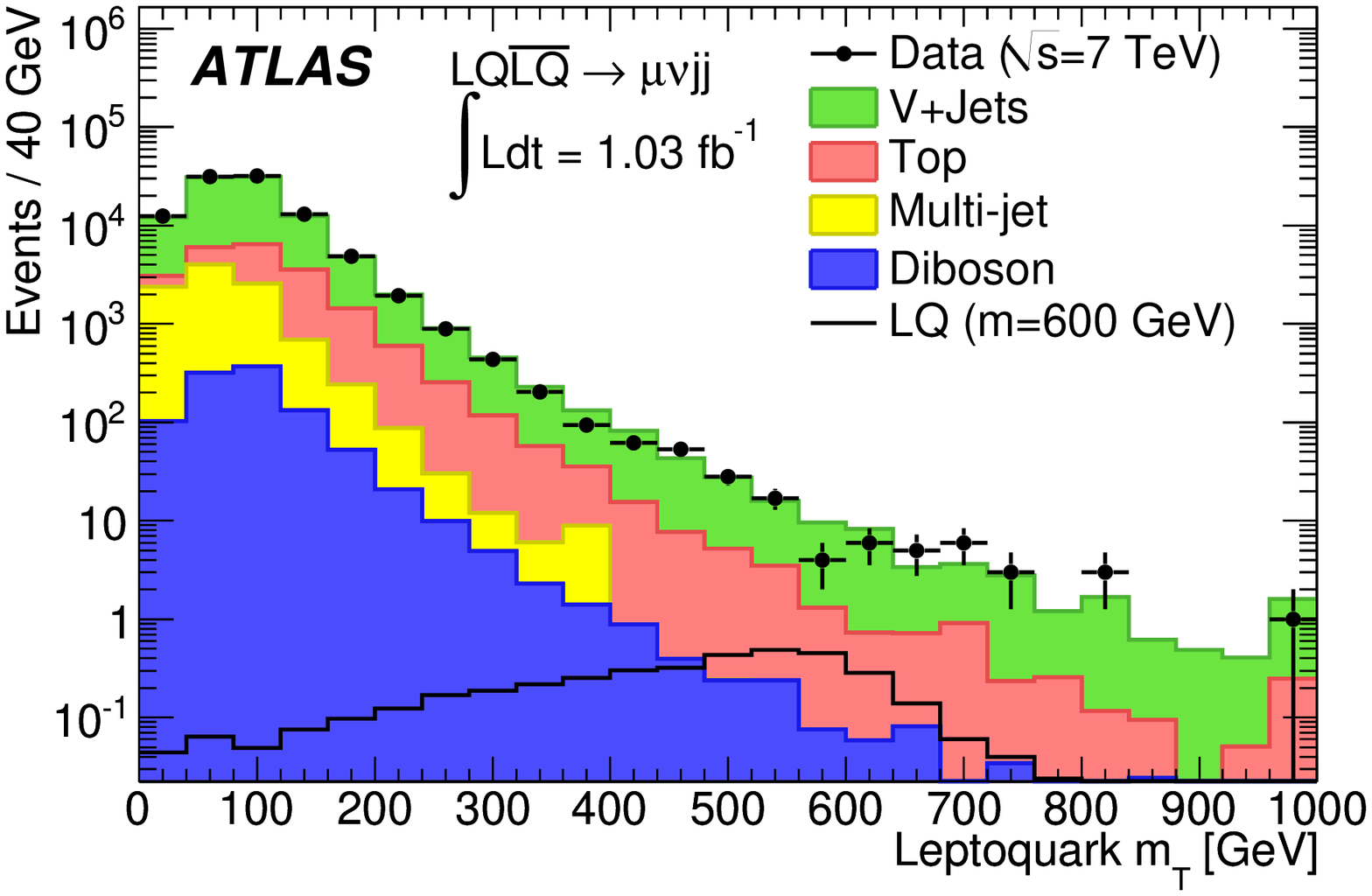}\label{f-lqmt}}
  \caption{Distributions of the input $LLR$ variables for the $\mnjj$ channel for data and the SM backgrounds.
 \subref{f-mt} Transverse mass of the muon and the $\MET$ in the event,
  \subref{f-st-mvjj} $S_{\rm T}$, \subref{f-lqm-mvjj} LQ mass, and \subref{f-lqmt} LQ transverse mass.
  The stacked distributions show the various background
  contributions, and data are indicated by the points with error bars. The expected signal for a 600~GeV LQ
 signal is also shown for $\beta=0.5$.  In all figures, the last bin contains the sum of all entries equal to and above the bin 
    lower boundary.\label{f-inputs-mvjj}}
\end{figure*}

\section{Background determination}
\label{sec-bkg}

Major backgrounds in this search arise from $V$+jets ($V=W,Z$) and $\ttbar$ processes.
The kinematic distributions of 
these are determined using MC samples, and their absolute normalization is evaluated from data 
using control regions, which are subsets of the selected sample, designed to enhance either 
the $V$+jets or the top quark contribution. 
The multi-jet background is obtained directly from data and prior to the estimation of the normalization
for the two main backgrounds, while the determination of the 
remaining backgrounds (diboson and single top quark production) relies entirely on MC simulations.

Two control regions are used in the $\mmjj$ channel. (I) $Z$+jets: formed by 
events within a narrow dimuon invariant mass $m_{\mu\mu}$ window around the $Z$ boson mass, defined 
by $81 < m_{\mu\mu} < 101$~GeV, and at least two jets, and (II) $\ttbar$: one of the muons is 
replaced by an electron resulting in events with a muon and an electron, and at least two jets.

Three control regions are used in the $\mnjj$ channel. (I) $W$+ 2 jets: events in the vicinity
of the $W$ boson Jacobian peak, selected by requiring $40 < m_{\rm T} < 120$~GeV, exactly two jets 
and $S_{\rm T} < 225$~GeV, where $S_{\rm T}$ is the scalar summed transverse energy $S_{\rm T}$, 
defined as $S_{\rm T}=p_{\rm T}^{\rm \mu}+\MET+p_{\rm T}^{\rm jet1}+p_{\rm T}^{\rm jet2}$, 
(II) $W$+ 3 jets: events passing the $40 < m_{\rm T} < 120$~GeV
requirement, with at least three jets and $S_{\rm T} < 225$~GeV, and (III) $\ttbar$:
events with at least four jets, with $p_{\rm T}^{\rm jet1}>50$~GeV and $p_{\rm T}^{\rm jet2}>40$~GeV.
In all of the control regions the expected signal yields are negligible.

The normalizations of the $V$+jets and $\ttbar$ backgrounds are obtained by comparing
data and MC yields in the control samples defined above. In the $\mmjj$ channel, each correction
factor is obtained independently for each background, on account of the high purity of the two different
control regions. In the $\mnjj$ channel, there is significant cross-region contamination
and therefore the number of $V$+jets and $t\bar t$ events is determined by simultaneously 
minimizing the $\chi^2$ formed by the differences between the observed and predicted SM yields 
in the three control regions. The resulting scale factors are of the order of 10\% in the low
$S_{\rm T}$ region.

The multi-jet background in the selected sample and in each control sample is obtained from a fit to 
the $m_{\mu\mu}$ and $\MET$ distribution in the $\mmjj$ and $\mnjj$ channels, 
respectively. In these fits, the relative fraction
of the multi-jet background is a free parameter, and the sum of the total predicted events is
constrained to be equal to the total observed number of events. The $V$+jets and
$\ttbar$ normalizations are not fixed. Multi-jet background arises predominantly from muons from 
secondary decays. Therefore, templates for the multi-jet background 
distributions are constructed from multi-jet enhanced samples of data events in which the muons 
fail the requirement on the transverse impact parameter or the isolation selection requirements described in 
Section~\ref{sec-objsel}.
In the $\mmjj$ channel, the $W$+jets contribution is estimated together with the 
multi-jet background. 
During this procedure, the $V$+jets and $\ttbar$ normalizations are fitted as well, providing an independent
estimate. The resulting values agree with those obtained from the control regions, which are the ones used in the analysis.

After analyzing 1.03 fb$^{-1}$ of data and applying the analysis requirements described in 
Section~\ref{sec-objsel}, good agreement is observed between the data and the SM expectation. The
observed and expected yields in the selected sample are 9254 and $9300 \pm 1700$ for the
$\mmjj$ channel, and 97113 and $97000 \pm 19000$ for the $\mnjj$ channel. For a LQ mass of
600~GeV, $8.2 \pm 0.4$ and $3.9 \pm 0.2$ events are expected for the $\mmjj$ and the
$\mnjj$ final states, respectively. The aforementioned uncertainites fully account for 
(the dominant) systematic and statistical uncertainties.

\section{Likelihood analysis}
\label{sec-llr}

Several kinematic variables, selected to provide the best discrimination between LQ events and SM backgrounds,
are combined in a log likelihood ratio in order to search for a LQ signal.
In the $\mmjj$ channel, $m_{\mu\mu}$, $S_{\rm T}=p_{\rm T}^{\rm \mu 1}+p_{\rm T}^{\rm \mu 2}+p_{\rm T}^{\rm jet1}+p_{\rm T}^{\rm jet2}$ and 
the average reconstructed leptoquark mass $\bar m_{\rm LQ}$ are used. In the $\mu\nu jj$ channel, $S_{\rm T}$,
$m_{\rm T}$, the transverse leptoquark mass $m_{\rm T}^{\rm LQ}$ and 
the leptoquark mass $m_{\rm LQ}$ are used. The distributions of these input variables are shown in 
Fig.~\ref{f-inputs-mmjj} and Fig.~\ref{f-inputs-mvjj} for the $\mmjj$ and the $\mnjj$ final 
states, respectively.

In the $\mmjj$ channel, an average LQ mass $\bar m_{\rm LQ}$ is defined for each event by 
reconstructing all possible combinations of lepton-jet pairs, using the two highest $p_{\rm T}$ 
jets in each event. Of the four possible combinations in each event, the pairing which provides 
the smallest difference between the LQ masses is chosen, and their average is used in the
likelihood analysis. In the $\mnjj$ final state, because the longitudinal component of the neutrino 
momentum is unknown, only one mass from the muon and a jet can be reconstructed, and the $\MET$ and
the remaining jet are used to calculate the transverse mass of the other LQ. The two masses which
provide the smallest absolute difference are used in the likelihood analysis. With this algorithm,
the probability of picking the correct pairing is of around 90\% for both channels.

For each event, likelihoods are constructed for the background ($L_B$) and the various signal LQ
hypothesis ($L_S$) as follows: $L_B \equiv \prod b_i(x_{ij})$, $L_S \equiv \prod s_i(x_{ij})$, 
where $b_i$, $s_i$ are the probabilities of the $i$--th input variable from the normalized 
summed background and signal distributions, respectively, and $x_{ij}$ is the value of that variable 
for the $j$--th event in a sample. The log likelihood ratio for each tested signal,
$LLR=$log$(L_S/L_B)$, is used as the final variable to search for the LQ signal.

\begin{table}
\begin{center}
\caption{The predicted and observed yields and the expected yields for a LQ signal of
$m_{\rm LQ}=600$~GeV after requiring $LLR\ge 2$ for the $\mmjj$ channel and $LLR\ge 7$ for 
the $\mnjj$ channel. The $\mmjj$ ($\mnjj$) channel 
signal yields are computed assuming $\beta=1.0\,\,(0.5)$. Statistical and 
systematic uncertainties as described in Section~\ref{sec-sys} are shown.
These are calculated assuming a 100\% correlation
for the same source between the different backgrounds. These systematic uncertainties are computed 
as the sum of the absolute values of the systematic variation in each bin and are shown to indicate the 
scale.  This is an approximation to the standard ensemble method used in the limit setting code.
    \label{t-yields}}
  \begin{tabular}{lcc}\hline
   Source  & $\mmjj$ Channel & $\mnjj$ Channel\\
\hline\hline
$V$+jets &  $14.2 \pm 6.4$  & $12.9\pm 9.9$    \\
Top      &      $3.0 \pm 2.2$   & $1.9 \pm 1.2$    \\
Diboson  &  $0.8 \pm 0.6$   & $0.3 \pm 0.1$  \\
Multi-jet&    $<0.1$          & $<0.1$      \\ \hline\hline
Total    &     $18 \pm 8$      & $15 \pm 11$   \\
Data     &     16              & 14            \\ \hline
LQ       &      $8.2 \pm 0.4 $  & $3.2 \pm 0.2$  \\
 \hline
 \end{tabular}
\end{center}
\end{table}

\begin{figure}[!htbp]
\centering
  \subfigure[]{\includegraphics[width= 0.52\textwidth]{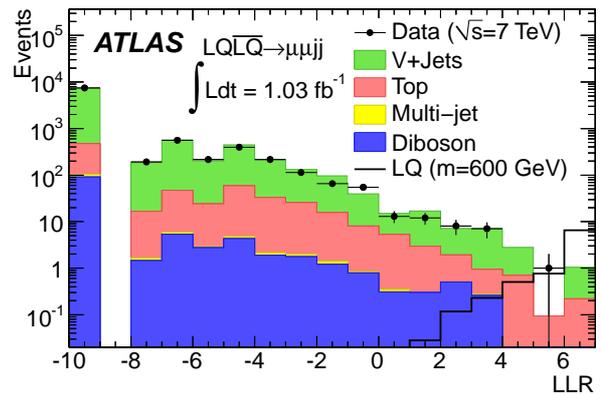}\label{f-llr-mmjj}}
   \subfigure[]{\includegraphics[width= 0.47\textwidth]{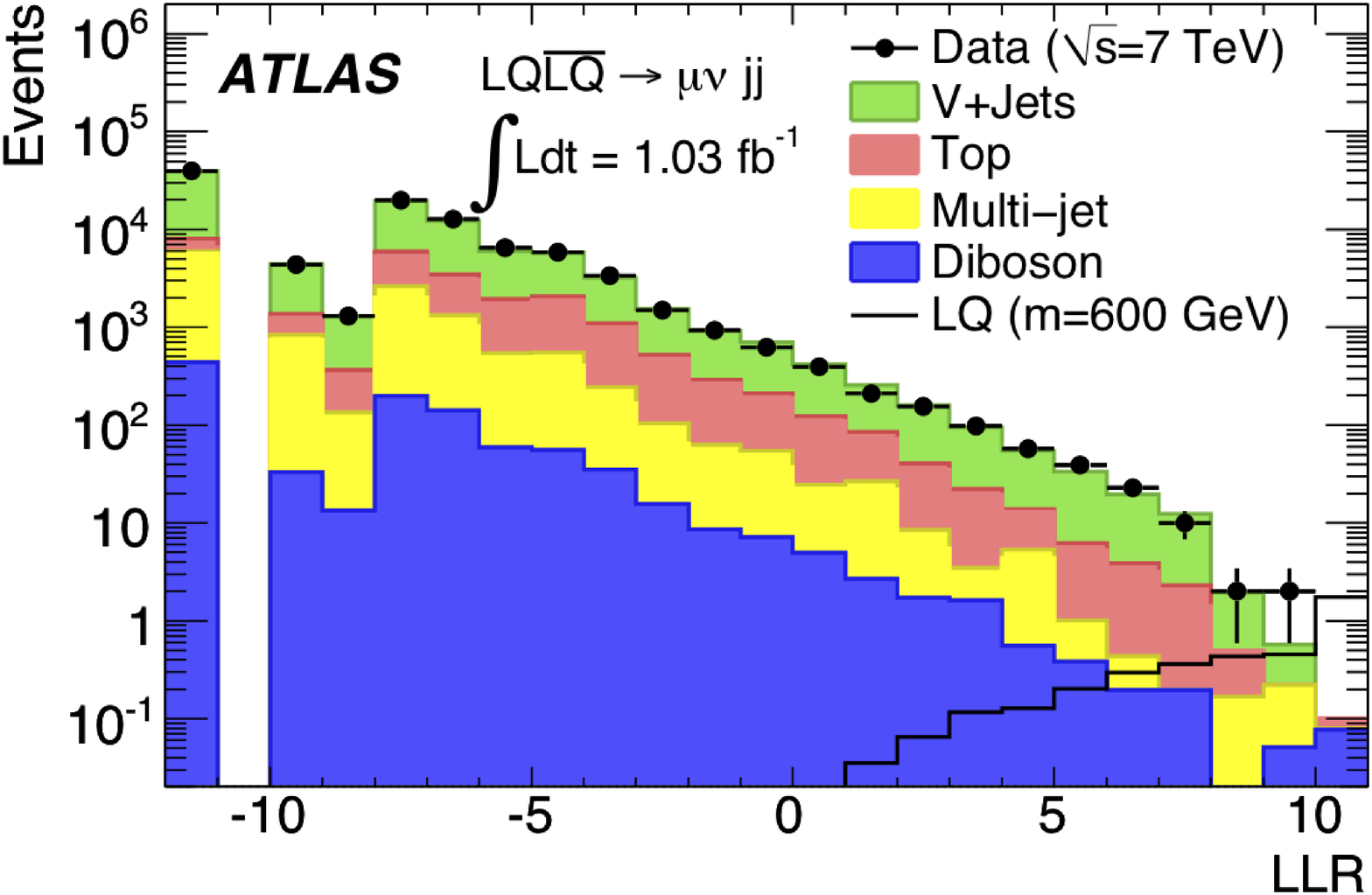}\label{f-llr-mvjj}}
    \caption{\subref{f-llr-mmjj} $LLR$ distributions for the $\mmjj$  
  and \subref{f-llr-mvjj} for the $\mnjj$ final states for a LQ mass of 600~GeV.
  The data are indicated with the points and the filled histograms show the SM background.
  The multi-jet background is estimated from data, while the other background contributions
  are obtained from simulated samples as described in the text. The LQ signal corresponding to a LQ mass of 600~GeV
  is indicated by a solid line, and is normalized assuming $\beta=1.0(0.5)$ in the $\mmjj$ ($\mnjj$) channel. 
  The lowest bin corresponds to background events in regions of
  the phase space for which no signal events are expected. 
   \label{f-LLR}}
\end{figure}

\section{Systematic uncertainties}
\label{sec-sys}

Systematic uncertainties originating from several sources are considered. 
These include uncertainties in lepton momentum, jet energy and $\MET$ scales
and resolutions and their dependence on the number of pile-up events, the background
estimations, and the LQ production cross section. For each source of
uncertainty considered, the analysis is repeated with the relevant variable
varied within its uncertainty, and a new $LLR$ is built for the systematically varied sample,
enabling the uncertainty in both the predicted yield and the kinematic
distributions to be propagated to the final result. In this section, systematic uncertainties
are described for each source of systematics, calculated assuming each source to be 100\% correlated
among the different backgrounds. Uncertainties are given for the
region of $LLR\geq2$ and $LLR\geq7$ for the $\mmjj$ and the $\mnjj$ channels,
respectively, although the full $LLR$ distribution is used to search for the LQ signal. 

The jet energy scale (JES) and resolution (JER) are varied up and down by 1$\sigma$~\cite{b-jetpaper}
 for all simulated events. Their impact is estimated independently, and the corresponding
variations are propagated to the $\MET$ in the case of the $\mnjj$ channel. The resulting effect of the
JES (JER) uncertainty is 9\% (8\%) and 15\% (7\%) for the backgrounds in the $\mmjj$ 
and the $\mnjj$ channels, respectively. For a LQ signal of $m_{\rm LQ}=600$~GeV, both are 1\% for the $\mmjj$
channel, and 2.4\% and 1\% for the $\mnjj$ channel.

The systematic uncertainties from the muon resolution and momentum scale are derived by  
comparing the $m_{\mu\mu}$ distribution in $Z\to \mu \mu$ control samples to 
$Z\to\mu\mu$ MC samples and are approximately 1\%~\cite{b-smear}. These result in uncertainties of 12\% and 
3\% for the total background
prediction in the $\mmjj$ and the $\mnjj$ channels, respectively, and in uncertainties
of 1.4\% for a LQ signal of $m_{\rm LQ}=600$~GeV for the $\mmjj$ and the $\mnjj$ channels.

Systematic uncertainties due to assumptions in the modelling of the $V$+jets background
are estimated by using SHERPA~\cite{b-sherpa} samples instead of the ALPGEN samples described
in Section~\ref{sec-mc}. The resulting uncertainty is 30\% for the $\mmjj$ channel and 60\% 
for the $\mnjj$ channel. 
Similarly, systematic uncertainties arising from the modelling of 
the $t\bar t$ process are
obtained by using different parameter values to simulate alternative samples to the one described in Section~\ref{sec-mc}. 
These include samples
in which the top quark mass is varied up and down by 2.5~GeV, generated with MC@NLO, samples 
where the initial and final-state radiation 
(ISR and FSR) contributions are varied accordingly to their uncertainties, generated with
ACER MC~\cite{b-acer}, and samples generated with POWHEG~\cite{b-powheg} interfaced to PYTHIA 
and JIMMY. These impact the total background yields by 12\% (7\%) for the $\mmjj$ ($\mnjj$) 
final state. For both $V$+jets and $\ttbar$ backgrounds, a 10\% uncertainty on the scale factors
is considered, covering the variation of the scale factors in the low and high $p_{\rm T}$ regions.

Systematic uncertainties in the multi-jet background in the $\mmjj$ channel are determined by 
comparing results derived from fits to kinematic variables other than the nominal ones. 
These include the
leading muon $p_{\rm T}$, the leading jet $p_{\rm T}$, the $\MET$ and the scalar sum
of the transverse momenta of the two muons in the event. In the $\mnjj$ channel, an 
alternative loose-tight matrix method~\cite{b-abcd} with two different multi-jet enhanced 
samples obtained by inverting the isolation and the $|d_0|$ requirements is used. 
Since the relevant phase space of the multi-jets in the two channels is very different, the
different control regions have very different statistics which leads to a large difference in 
precision to which this background can be estimated. The resulting
uncertainties are 90\% in the $\mmjj$ channel and 33\% in the $\mnjj$ channel.

A luminosity uncertainty of 3.7\%~\cite{b-lumi} is assigned to the LQ signal yields
and to the yields of background processes determined from simulation: diboson and single top
quark production.
Further systematic uncertainties considered arise from
the finite number of events in the simulated samples, amounting to 4\%--25\% 
depending on the LQ mass being considered.

For the signal samples, additional systematic uncertainties originate from ISR and FSR
effects, resulting in an uncertainty of 2\% for both channels. The choice of the 
renormalization and factorization scales, which are varied from $m_{\rm LQ}$
to 2$m_{\rm LQ}$ and $m_{\rm LQ}/2$, and the choice of the PDF, 
determined with the CTEQ eigenvectors errors
and by using the MRST2007LO* PDF set~\cite{b-sysPDF}, result 
in an uncertainty in the signal acceptance of 1\%--6\% for LQ masses between 300~GeV and 700~GeV.

\section{Results}
\label{sec-res}

Figure~\ref{f-LLR} shows the $LLR$ for the data, the predicted backgrounds and a LQ signal of 600~GeV 
for the $\mmjj$ and the $\mnjj$ channels. To ensure sufficient background statistics, bins
with a total background yield less than twice the statistical uncertainty in that bin are
merged into a single bin. There is no 
significant excess in data observed at large $LLR$ values where such a signal would appear, 
and the data are found to be consistent with the SM background expectations (see Table~\ref{t-yields}). 
Upper limits are derived at 95\% confidence level (CL) for the scalar leptoquark production 
cross section using a modified 
frequentist $CL_{s}$ approach~\cite{b-lcalc1,b-lcalc2}. The test statistic is defined as 
$-2\ln(Q) = -2 \ln(L_{s+b} / L_{b} ) $, where the likelihoods $L_{s+b}$ and $L_{s}$ follow
a Poisson distribution and are calculated based on the corresponding $LLR$ distributions. 
Systematic uncertainties as 
described in Section~\ref{sec-sys} are treated as nuisance parameters with a Gaussian probability 
density function.

\begin{figure}[!htbp]
\centering
  \subfigure[]{\includegraphics[width= 0.49\textwidth]{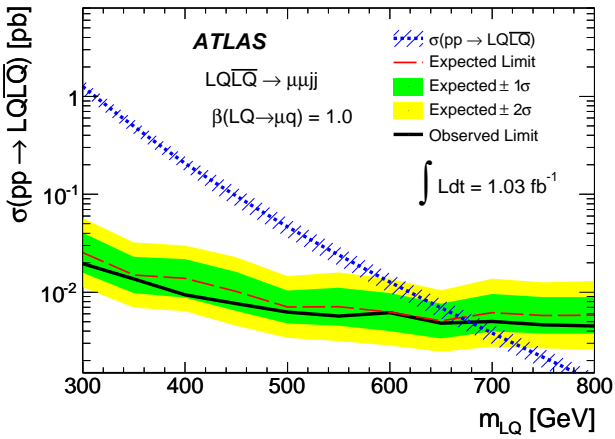}\label{f-mmjj}}
  \subfigure[]{\includegraphics[width= 0.49\textwidth]{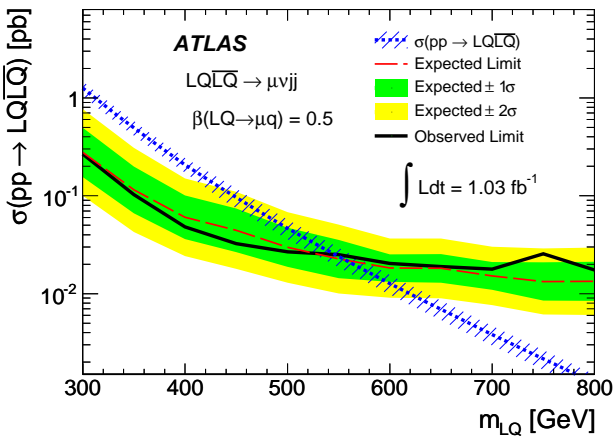}\label{f-mvjj}}
    \caption{\subref{f-mmjj} 95\% CL upper limit on  the pair production cross section 
    of second generation leptoquarks for the $\mmjj$ channel at $\beta=$1.0 
    and \subref{f-mvjj} for the $\mnjj$ channel at $\beta=$0.5. The solid lines indicate the individual
    observed limits, while the expected limits are indicated by the dashed lines. The theoretical
    prediction is indicated by the hatched band and includes the systematic uncertainties due to
    the choices of the PDF and the renormalization and factorization scales.
    The dark (green) and light (yellow) solid band contains 68\% (95\%), respectively, of possible outcomes
    from pseudo-experiments in which the yield is Poisson-fluctuated around the background-only expectation.
\label{f-xsec95}}
\end{figure}

\begin{figure}[!htbp]
  \includegraphics[width=0.97\columnwidth]{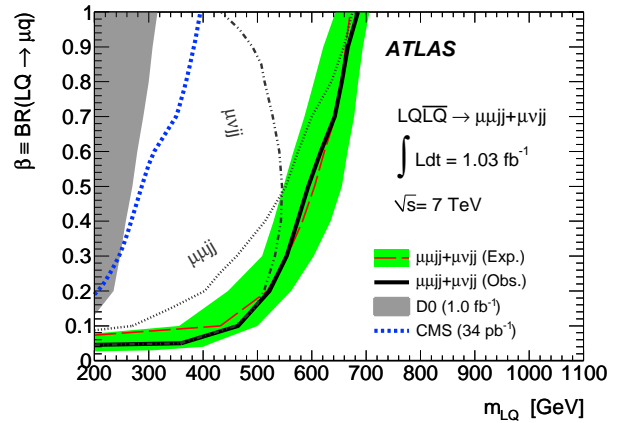}
  \caption{95\% CL exclusion region resulting from the combination of the
  $\mmjj$ and the $\mnjj$ channels shown in the $\beta$ versus leptoquark mass plane.
  The shaded area at the left indicates the D0 exclusion limit~\cite{LQ_Fermilab} and the thick dotted line
indicates the CMS exclusion region~\cite{LQ_LHC1}. The dotted and dotted-dashed lines
  indicate the individual limits derived for the $\mmjj$ and $\mnjj$ channels, respectively. The combined observed limit is indicated by 
the solid black line. The combined expected limit is indicated by the dashed line, together with the solid band containing 68\% of 
possible outcomes from pseudo-experiments in which the yield is Poisson-fluctuated around the background-only expectation.
\label{f-bvm}}
\end{figure}

The 95\% CL upper bounds on the cross section for leptoquark pair production as 
a function of mass are shown in Fig.~\ref{f-xsec95} for the $\mmjj$ and the $\mnjj$ channels
at $\beta=1.0$ and $\beta=0.5$, respectively. The expected and observed limits for the combined channels are 
shown in the $\beta$ vs. $m_{\rm LQ}$ plane in Fig.~\ref{f-bvm}. 

\section{Conclusions}

The results of a search for the pair production of second generation scalar leptoquarks 
using 1.03~fb$^{-1}$ of proton-proton collision data produced by the LHC at $\sqrt{s}=7$~TeV
and recorded by the ATLAS detector are presented. The data are in good agreement with the expected
SM background, and no evidence of LQ production is observed. Lower limits 
on leptoquark masses of $m_{\rm LQ}>685$~GeV and $m_{\rm LQ}>594$~GeV for $\beta=1.0$ and $\beta=0.5$
are obtained at 95\% CL, whereas the expected limits are $m_{\rm LQ}>671$~GeV and $m_{\rm LQ}>605$~GeV, respectively. 
These are the most stringent limits to date arising from direct searches for second generation scalar leptoquarks.

\vspace{0.2cm}

\section{Acknowledgements}

We thank CERN for the very successful operation of the LHC, as well as the
support staff from our institutions without whom ATLAS could not be
operated efficiently.

We acknowledge the support of ANPCyT, Argentina; YerPhI, Armenia; ARC,
Australia; BMWF, Austria; ANAS, Azerbaijan; SSTC, Belarus; CNPq and FAPESP,
Brazil; NSERC, NRC and CFI, Canada; CERN; CONICYT, Chile; CAS, MOST and NSFC,
China; COLCIENCIAS, Colombia; MSMT CR, MPO CR and VSC CR, Czech Republic;
DNRF, DNSRC and Lundbeck Foundation, Denmark; ARTEMIS and ERC, European Union;
IN2P3-CNRS, CEA-DSM/IRFU, France; GNAS, Georgia; BMBF, DFG, HGF, MPG and AvH
Foundation, Germany; GSRT, Greece; ISF, MINERVA, GIF, DIP and Benoziyo Center,
Israel; INFN, Italy; MEXT and JSPS, Japan; CNRST, Morocco; FOM and NWO,
Netherlands; RCN, Norway; MNiSW, Poland; GRICES and FCT, Portugal; MERYS
(MECTS), Romania; MES of Russia and ROSATOM, Russian Federation; JINR; MSTD,
Serbia; MSSR, Slovakia; ARRS and MVZT, Slovenia; DST/NRF, South Africa;
MICINN, Spain; SRC and Wallenberg Foundation, Sweden; SER, SNSF and Cantons of
Bern and Geneva, Switzerland; NSC, Taiwan; TAEK, Turkey; STFC, the Royal
Society and Leverhulme Trust, United Kingdom; DOE and NSF, United States of
America.

The crucial computing support from all WLCG partners is acknowledged
gratefully, in particular from CERN and the ATLAS Tier-1 facilities at
TRIUMF (Canada), NDGF (Denmark, Norway, Sweden), CC-IN2P3 (France),
KIT/GridKA (Germany), INFN-CNAF (Italy), NL-T1 \\(Netherlands), PIC (Spain),
ASGC (Taiwan), RAL (UK) and BNL (USA) and in the Tier-2 facilities
worldwide.

\onecolumn
\clearpage
\begin{flushleft}
{\Large The ATLAS Collaboration}

\bigskip

G.~Aad$^{\rm 48}$,
B.~Abbott$^{\rm 112}$,
J.~Abdallah$^{\rm 11}$,
S.~Abdel~Khalek$^{\rm 116}$,
A.A.~Abdelalim$^{\rm 49}$,
A.~Abdesselam$^{\rm 119}$,
O.~Abdinov$^{\rm 10}$,
B.~Abi$^{\rm 113}$,
M.~Abolins$^{\rm 89}$,
O.S.~AbouZeid$^{\rm 159}$
% ...
\bigskip

% ...
$^{10}$ Institute of Physics, Azerbaijan Academy of Sciences, Baku, Azerbaijan\\
$^{11}$ Institut de F\'isica d'Altes Energies and Departament de F\'isica de la Universitat Aut\`onoma  de Barcelona and ICREA, Barcelona, Spain\\
% ...
$^{48}$ Fakult\"{a}t f\"{u}r Mathematik und Physik, Albert-Ludwigs-Universit\"{a}t, Freiburg i.Br., Germany\\
$^{49}$ Section de Physique, Universit\'e de Gen\`eve, Geneva, Switzerland\\
% ...
$^{89}$ Department of Physics and Astronomy, Michigan State University, East Lansing MI, United States of America\\
% ...
$^{112}$ Homer L. Dodge Department of Physics and Astronomy, University of Oklahoma, Norman OK, United States of America\\
$^{113}$ Department of Physics, Oklahoma State University, Stillwater OK, United States of America\\
% ...
$^{116}$ LAL, Univ. Paris-Sud and CNRS/IN2P3, Orsay, France\\
% ...
$^{119}$ Department of Physics, Oxford University, Oxford, United Kingdom\\
% ...
$^{159}$ Department of Physics, University of Toronto, Toronto ON, Canada\\
% ...
\end{flushleft}



\begin{thebibliography}{99}

\bibitem{lqpapers} W.~Buchm{\"u}ller and D.~Wyler, Phys. Lett. {\bf B177} (1986) 377; 
J.~C.~Pati and A.~Salam, Phys.~Rev. {\bf D10} (1974) 275;
H.~Georgi and S.~Glashow, Phys. Rev. Lett. {\bf 32} (1974) 438;
S.~Dimopoulus and L.~Susskind, Nucl.~Phys. {\bf B155} (1979), 237; 
S.~Dimopoulus, Nucl.~Phys. {\bf B168} (1980) 69; 
E.~Eichten and K.~Lane, Phys.~Lett. {\bf B90} (1980) 125; 
B.~Schrempp and F.~Schrempp, Phys. Lett. {\bf B153} (1985) 101;
V.~D.~Angelopoulus et al., Nucl.~Phys. {\bf B292} (1987) 59.

\bibitem{mBRW} W. Buchm{\"u}ller et al.,
Phys.~Lett. {\bf B191} (1987) 442.

\bibitem {b-comphep} A.~Belyaev, C.~Leroy, R.~Mehdiyev, A.~Pukhov, JHEP {\bf 09} (2005) 005.

\bibitem{b-kramer}  M.~Kramer, T.~Plehn, M.~Spira and P.~M.~Zerwas,
  Phys. Rev. {\bf D71} (2005)  057503.

\bibitem{LQ_LHC2} ATLAS Collaboration, Phys. Rev. {\bf D83} (2011) 112006.

\bibitem{LQ_LHC1} CMS Collaboration, Phys. Rev. Lett. {\bf 106} (2011) 201803.

\bibitem{LQ_Fermilab} D0 Collaboration, V.~M.~Abazov et al., 
Phys. Lett. {\bf B636} (2006) 183.

\bibitem{LQ_LEP} OPAL Collaboration, G.~Abbiendi et al., 
Eur. Phys. J. {\bf C31} (2003) 281.

\bibitem{atlas} ATLAS Collaboration, JINST {\bf 3}, S08003 (2008).

\bibitem{b-pythia} T.~Sjostrand, S.~Mrenna and P.~Skands, 
 JHEP {\bf 05} (2006) 026.

\bibitem{b-cteq66} D. Stump et al., 
JHEP {\bf 10} (2003)  046;

\bibitem{b-alpgen} M.~Mangano et al.,
 JHEP {\bf 07} (2003) 001.

\bibitem{b-herwig} G. Corcella et al., 
 JHEP {\bf 01} (2001) 010.

\bibitem{b-jimmy} J.~Butterworth, J.~Forshaw and M.~Seymour,
 Z. Phys. {\bf C72} (1996) 637.

\bibitem{b-mcnlo} S.~Frixione and B.~R.~Webber,
JHEP {\bf 06} (2002) 029;
 S.~Frixione, P.~Nason and B.~R.~Webber, 
 JHEP {\bf 08} (2003) 007.

\bibitem{b-mcfm} J.~M.~Campbell and R.~K.~Ellis, 
Phys. Rev. {\bf D60}, 113006 (1999).

\bibitem{b-geant4} S. Agostinelli et al., GEANT4 Collaboration, 
Nucl. Instrum. Meth. {\bf A506} (2003) 250.
 
\bibitem{b-atlassim} ATLAS Collaboration, 
 Eur. Phys. J. {\bf C70} (2010) 823.

\bibitem{MCP} 
ATLAS Collaboration, Phys. Rev. {\bf D85} (2012) 072004.
\bibitem{b-smear} ATLAS Collaboration, ATLAS-CONF-2011-046, \\
{\tt http://cdsweb.cern.ch/record/1338575}.

\bibitem{b-akt}  M.~Cacciari, G.~P.~Salam and G.~Soyez,  JHEP {\bf 04}, (2008) 063;
 M.~Cacciari and G.~P.~Salam, Phys. Lett. {\bf B641} (2006) 57.

\bibitem{b-jets-testbeam}  P. Adragna et al., Nucl. Instrum.  Meth. {\bf A615}, (2010) 158.

\bibitem{b-jetpaper} ATLAS Collaboration, arXiv:1112.6426, Submitted to Eur. Phys. J. C.

\bibitem{b-lq1} ATLAS Collaboration, Phys. Lett. {\bf B709}, 158-176 (2012).

\bibitem{b-sherpa} T.~Gleisberg et al. JHEP {\bf 02} (2009) 007; 
S. Schumann and F. Krauss, JHEP {\bf 03} (2008) 038; 
S. Hoeche, F. Krauss, S. Schumann, and F. Siegert, JHEP {\bf 05} (2009) 053.

\bibitem{b-acer} B.~P.~Kersevan and E.~Richter-Was, arXiv:0405247 (2004).

\bibitem{b-powheg} S.~Frixione, P.~Nason and C.~Oleari, 
JHEP {\bf 11}(2007) 070.

\bibitem{b-abcd} ATLAS Collaboration, Eur. Phys. J. {\bf C71}, 1577 (2011).

\bibitem{b-lumi} ATLAS Collaboration, ATLAS-CONF-2011-116, \\
{\tt http://cdsweb.cern.ch/record/1376384}; 

ATLAS Collaboration, Eur. Phys. J {\bf C71}, 1630 (2011).

\bibitem{b-sysPDF} A.~Sherstnev and R.~S.~Thorne, Eur. Phys. J. {\bf C55}, 553 (2008).

\bibitem{b-lcalc1}
 W.~Fisher, 
 FERMILAB-TM-2386-E.

\bibitem{b-lcalc2}
 T.~Junk,
  Nucl. Instrum. Meth. {\bf A434}, (1999) 435.

\end{thebibliography}
\end{document}